\documentclass[journal]{IEEEtran}

\usepackage{amsmath,amssymb,amsfonts}
\usepackage{algorithmic}
\usepackage{graphicx}

\usepackage{multirow}
\usepackage{subcaption}
\usepackage{caption}
\usepackage{float}
\usepackage{xcolor}
\usepackage{booktabs}
\usepackage{hyperref}

\usepackage[switch]{lineno}

\begin{document}


%

\title{Limitations of Out-of-Distribution Detection in 3D Medical Image Segmentation}

\author{
Anton Vasiliuk,  
Daria Frolova,  
Mikhail Belyaev,  
and Boris Shirokikh  

\thanks{The results have been obtained under the support of the Russian Science Foundation grant number 20-71-10134. (Corresponding author: B. Shirokikh.)}

\thanks{A. Vasiliuk is 
with the Moscow Institute of Physics and Technology, Dolgoprudny, Moscow Region, 141701, Russia
and Artificial Intelligence Research Institute (AIRI), Moscow, 105064, Russia
(e-mail: vasilyuk@phystech.edu).}

\thanks{D. Frolova, M. Belyaev, and B. Shirokikh are
with the Skolkovo Institute of Science and Technology, Moscow, 143026, Russia
and Artificial Intelligence Research Institute (AIRI), Moscow, 105064, Russia (e-mail: frolova@airi.net; m.belyaev@skoltech.ru; boris.shirokikh@skoltech.ru).}

\thanks{\textit{This work has been submitted to the IEEE for possible publication. Copyright may be transferred without notice, after which this version may no longer be accessible}.}
}

\maketitle

\begin{abstract}
Deep Learning models perform unreliably when the data comes from a distribution different from the training one. In critical applications such as medical imaging, out-of-distribution (OOD) detection methods help to identify such data samples, preventing erroneous predictions. In this paper, we further investigate the OOD detection effectiveness when applied to 3D medical image segmentation. We design several OOD challenges representing clinically occurring cases and show that none of these methods achieve acceptable performance. Methods not dedicated to segmentation severely fail to perform in the designed setups; their best mean false positive rate at 95\% true positive rate (FPR) is 0.59. Segmentation-dedicated ones still achieve suboptimal performance, with the best mean FPR of 0.31 (lower is better). To indicate this suboptimality, we develop a simple method called Intensity Histogram Features (IHF), which performs comparable or better in the same challenges, with a mean FPR of 0.25. Our findings highlight the limitations of the existing OOD detection methods on 3D medical images and present a promising avenue for improving them. To facilitate research in this area, we release the designed challenges as a publicly available benchmark and formulate practical criteria to test the OOD detection generalization beyond the suggested benchmark. We also propose IHF as a solid baseline to contest the emerging methods.
\end{abstract}



\section{Introduction}
\label{sec:intro}

In recent years, Deep Learning (DL) methods have achieved human-level performance in automated medical image processing. But the development of these methods on a large scale is slowed by several factors. One such factor is the unreliable performance of DL models when the data comes from a distribution different from the training one \cite{wang2018deep}. These differences are common in medical imaging: population, demographic, acquisition parameter changes, or new imaging modalities.

Out-of-distribution (OOD) detection helps to identify the data samples with such differences, hence increasing the reliability and safety of a DL model. For instance, detected cases could be marked as rejected, preserving the model performance, or reported to the experts, preventing the model from failing silently. The ability to report or reject unreliable cases is now considered a necessary capability to enable safe clinical deployment \cite{kompa2021second}.

OOD detection on natural images is a well-researched area \cite{yang2021generalized} where several established benchmarks \cite{hendrycks2016baseline,hendrycks2019scaling} facilitate its development. Moreover, these methods directly scale on 2D medical images, resulting in multiple algorithms \cite{mahmood2020multiscale,pacheco2020out,berger2021confidence}, and also a benchmark \cite{cao2020benchmark}. At the same time, OOD detection on 3D medical images remains poorly explored, although 3D medical image segmentation is one of the most addressed tasks in medical imaging \cite{litjens2017survey} with outstanding practical usefulness, e.g., quantifying anatomical structures, pathologies, or important biomarkers.

The primary cause of this poor exploration is the lack of datasets and benchmarks with a correct problem design. For example, one party uses private data \cite{karimi2022improving}, while the other simulates synthetic anomalies that are unlikely to occur in clinical settings \cite{david_zimmerer_2022_6362313}. A study can be limited to a single distribution shift, e.g., changes in the scanning location \cite{karimi2022improving}, lacking the diversity of setups. Also, a study can be restricted to uncertainty estimation \cite{lambert2022improving} or anomaly detection \cite{david_zimmerer_2022_6362313} methods, leaving the full spectrum of approaches uncovered. Such issues limit a fair comparison of the proposed approaches.

In this paper, we investigate the effectiveness of OOD detection when applied to 3D medical image segmentation, closing the outlined gaps in prior work. To enable a correct comparison, we thus design a \textit{diverse} set of challenges using \textit{publicly available data} with a \textit{downstream segmentation task} and simulating \textit{clinically occurring anomaly sources}. Besides the problem design, such a study requires appropriately selected state-of-the-art methods. We note that several areas, e.g., anomaly detection and uncertainty estimation, share motivation and methodology with OOD detection. Therefore, we overview all related areas and, contrary to the previous works, present a complete methodological coverage.

Extensive evaluation of six selected methods results in our main conclusion: state-of-the-art OOD detection falls short of achieving optimal performance on 3D medical images. We show that the methods not designed for segmentation completely fail in most setups, scoring from $0.84$ to $0.59$ False-Positive Rate (FPR) on average, which is not far below $0.95$ FPR of the random guessing. (Lower FPR is better.) Two methods specifically designed for 3D segmentation achieve $0.38$ and $0.31$ mean FPR, further reducing the error about two times. At the same time, we show that these errors can be reduced even further with a simple approach.

We show this space for improvement by developing a histogram-based method called Intensity Histogram Features (IHF). IHF achieves comparable and often superior results to its competitors with $0.25$ mean FPR. It also scores $0$ FPR in multiple challenges, indicating that the distribution shifts in 3D medical imaging can often be detected using image intensity histograms, while the DL-based methods overlook this domain feature. Therefore, we consider current DL-based OOD detection far from unveiling its full potential and assume it can be further improved.

Given IHF's negligible computational costs compared to DL, we suggest it as a baseline to contest the emerging OOD detection methods. Furthermore, we propose using the designed challenges as a benchmark for developing new methods. Correct problem setting, in-depth analysis with simple methods, such as IHF, and ablation studies on synthetic data confirm that our benchmark allows to estimate the quality of solving general OOD detection instead of classifying a priori known anomaly types. Thus, summarizing our contributions, we outline the following:

\begin{enumerate}

    \item We demonstrate severe limitations of the existing OOD detection methods on 3D medical images.

    \item We design and release the corresponding benchmark that can be used as a starting point for the related research. 

    
    \item We propose a method, IHF, suggesting it as a solid baseline for OOD detection on 3D medical images.
    
\end{enumerate}

Below, we describe the data used in our study and the problem setup (Sec.~\ref{sec:data}). Then, we review and select state-of-the-art and core methods from the related fields and also detail IHF (Sec.~\ref{sec:methods}). Finally, we present the results (Sec.~\ref{sec:results}) and discuss the limitations and implications of our study (Sec.~\ref{sec:discussion}).

\section{Data}
\label{sec:data}


\begin{figure*}[h]
    \begin{center}
        \includegraphics[width=\textwidth]{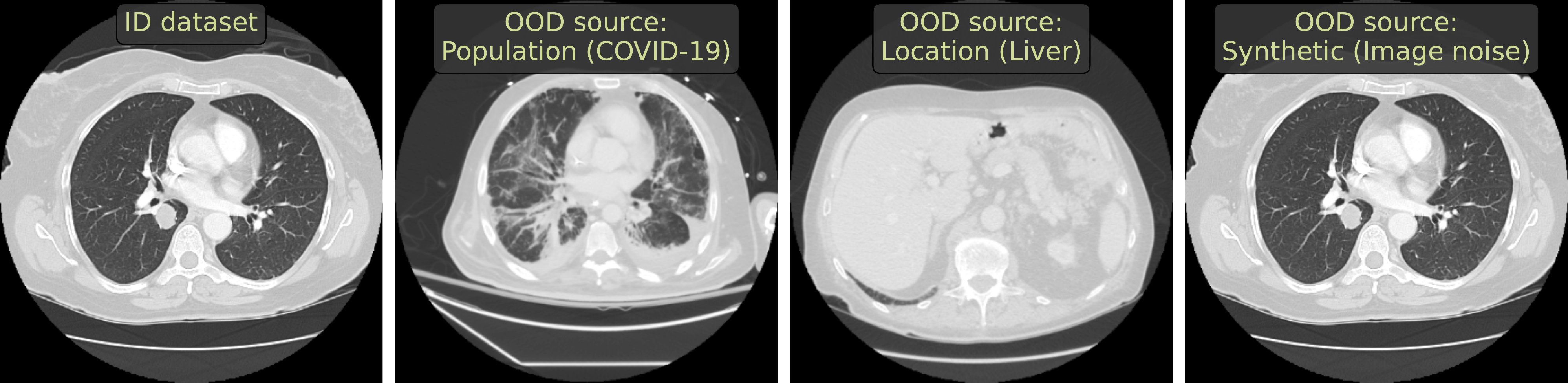}
        \caption{Examples of CT images (representative axial slices) from different simulated OOD sources in our benchmark.}
        \label{fig:ct}
    \end{center}
\end{figure*}

\begin{figure*}[h]
    \begin{center}
        \includegraphics[width=\textwidth]{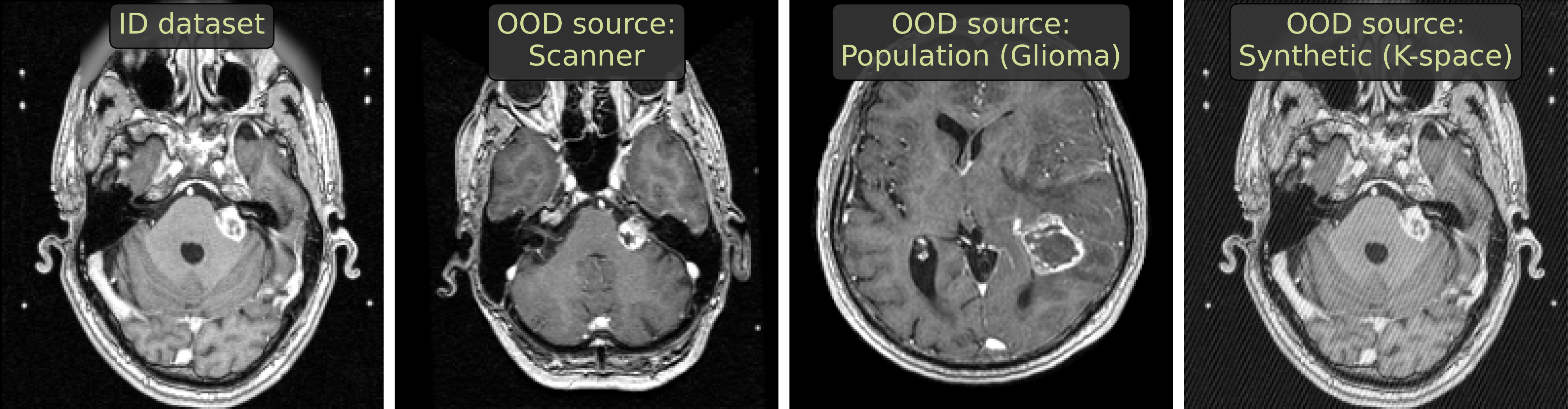}
        \caption{Examples of MRI images (representative axial slices) from different simulated OOD sources in our benchmark.}
        \label{fig:mri}
    \end{center}
\end{figure*}

Contrary to the fields of 2D natural and medical images, no established OOD detection benchmark with a correct problem setting exists for 3D medical images. For example, \cite{karimi2022improving} used a variety of brain and abdominal CT and MRI datasets but included private ones. The authors also studied only a single distribution shift, changes in the scanning location, which does not allow to estimate the general performance. \cite{david_zimmerer_2022_6362313} created an OOD detection challenge, simulating synthetic anomalies in brain MR and abdominal CT images. However, their setup lacks a downstream task (e.g., segmentation), so their study is limited to unsupervised anomaly detection methods. Synthesizing local corruptions, as in \cite{david_zimmerer_2022_6362313}, can also lead to evaluation biases, which we show with our analysis. On the other hand, \cite{lambert2022improving} included datasets with the segmentation task but limited the considered methods to supervised uncertainty estimation. 

Given the disagreement of setups, their partial problem coverage, or privacy, we design the OOD detection challenges from scratch following three core principles:

\begin{itemize}
    
    \item We include two large \textit{publicly available} CT and MRI in-distribution (ID) datasets to cover the most frequent volumetric modalities.
    
    \item We ensure both datasets have \textit{a downstream segmentation task}, allowing us to use the full spectrum of methods. 
    
    \item We select \textit{diverse} OOD datasets that simulate the \textit{clinically occurring sources of anomalies}: changes in acquisition protocol, patient population, or anatomical region. All these datasets are also publicly available.
    
\end{itemize}

We also introduce several medical imaging artifacts as anomaly sources, synthetically generated as random transformations. Generating synthetic anomalies is a popular approach, applied to 3D images \cite{david_zimmerer_2022_6362313,lambert2022improving} as well as 2D ones \cite{hendrycks2016baseline,hendrycks2019scaling}; this approach also allows us to conduct controlled ablation studies at different distortion levels.


We made the resulting benchmark publicly available\footnote{\url{https://github.com/francisso/OOD-benchmark}} and we detail all related CT and MRI datasets in Sec.~\ref{ssec:data:ct} and Sec.~\ref{ssec:data:mri}. The problem setting is described in Sec.~\ref{ssec:data:exp}.

\subsection{3D CT datasets}
\label{ssec:data:ct}

We construct a total of 6 challenges on CT data, including two synthetic ones. We give a visual example of data samples in Fig.~\ref{fig:ct} and detail the ID dataset and every setup below.

\subsubsection*{ID dataset}
As an ID CT dataset, we use LIDC-IDRI \cite{armato2011lung}. It contains $1018$ chest CT images with the lung nodules segmentation task. We remove cases without nodules since they do not contribute to training a segmentation model. Then, we randomly split the rest $883$ images $4 : 1$ into the train and test, stratified by the number of nodules.

\subsubsection{OOD source: scanner}
To simulate a covariate shift, we select Cancer500 \cite{cancer500} that has the same downstream task as the ID dataset but is obtained with different scanners and acquisition protocols. It contains $979$ chest CT images. We exclude all images with low resolution (less than $64$ axial slices) and no annotated nodules, resulting in $841$ images left.

\subsubsection{OOD source: population}
To simulate a patient population shift, we use two datasets with the similar semantic content but different downstream task. These datasets are Medseg9\footnote{\url{https://radiopaedia.org/articles/covid-19-3}} and MIDRC \cite{tsai2021rsna}, containing 9 and 154 chest CT images, respectively, with COVID-19 cases. Excluding all non-COVID cases, the merged dataset has 120 images.

\subsubsection{OOD source: location (liver)}
To simulate a semantic shift, we select a dataset of the same modality that focuses on different body region. Here, we use LiTS \cite{bilic2019liver}, a dataset with $201$ abdominal CT images. 

\subsubsection{OOD source: location (head)}
Similarly, we include CT-ICH \cite{hssayeni2020computed}, a dataset with $75$ head CT images. 

\subsubsection{OOD source: synthetic (image noise)}
\label{sssec:data:ct:noise}
We simulate local image corruptions by applying damaging transformations to the testing cases of ID dataset. Applied to a randomly selected image crop, these transformations include blurring, changing contrast, or inserting noise.

\subsubsection{OOD source: synthetic (elastic)}
We simulate tissue anomalies by applying elastic transform of random severity. 

\subsection{3D MRI datasets}
\label{ssec:data:mri}

We construct a total of 7 challenges on MRI data, including four synthetic ones. We give a visual example of data samples in Fig.~\ref{fig:mri} and detail every setup below.

\subsubsection*{ID dataset}
As an ID MRI dataset, we use VS-Seg \cite{shapey2021segmentation}. It contains $242$ brain T1c MRIs with the vestibular schwannoma segmentation task. We remove cases with empty target mask and split the rest $239$ images $2:1$ into the train and test.

\subsubsection{OOD source: scanner}
To simulate a covariate shift, we select the data with the same semantic content and downstream task but is obtained with different scanners and acquisition protocols. Here, we choose CrossMoDA ETZ as a subset of the CrossMoDA 2022 Challenge dataset \cite{reuben_dorent_2022_6504722} with $105$ brain T1c MR images and use it without changes.

\subsubsection{OOD source: population (glioblastoma)}
To simulate a patient population shift, we use EGD \cite{van2021erasmus}, a dataset with 774 brain MRIs of four modalities (FLAIR, T1, T1c, T2) with a glioma segmentation task. We reduce a possible covariate shift by using only T1c modality from the Siemens Avanto 1.5T scanner, as in VS-Seg, resulting in 262 seleted images.

\subsubsection{OOD source: population (healthy)}
Additionally, we simulate a patient population shift with healthy cases instead of the changing pathology. To do so, we use the CC359 \cite{souza2018open} dataset with $359$ brain MR images of T1 modality. We note however that CC359 images differ in vendor, scanning protocol, and do not contain contrast enhancement, so this setup has a secondary OOD source, a covariate shift.

\subsubsection{OOD source: synthetic (K-space noise)}
We synthesize MR imaging artifact, known as Herringbone artifact, at different magnitudes. This results in the visible spikes across the whole image due to anomaly points in K-space.

\subsubsection{OOD source: synthetic (anisotropy)}
We synthesize wrong resolution by downsampling the image and upsampling it back along one randomly chosen axis.

\subsubsection{OOD source: synthetic (motion)}
We synthesize two types of MR imaging artifacts that can happen due to the patient motion. One is ghosting, which appears as shifted copies of the original image. Another exploits \textit{RandomMotion} simulation from the \texttt{torchIO} library \cite{torchio}.

\subsubsection{OOD source: synthetic (image noise)}
The same pipeline as for CT images; see Sec.~\ref{sssec:data:ct:noise}.

\subsection{Problem setting}
\label{ssec:data:exp}

We define the OOD detection problem as the classification between samples from a source distribution (ID) and abnormal samples from a novel different distribution (OOD). The core assumption is the abnormal sample distribution is unknown and cannot be computed in advance. Thus, we approximate the anomaly distribution by constructing a diverse set of challenges that represent clinically occurring cases. Consequently, to develop a reliable method in practice, we need not only to attain the desired accuracy on such a set, but also to ensure that the method is capable of generalizing to novel sources of anomalies.




By providing a downstream segmentation task, we remove any constraints on the method design. One can use segmentation model features, its uncertainty estimates, or an auxiliary model to detect outliers. Trained algorithms output a single number called \textit{OOD score} for every testing image; a higher score means a higher outlier likelihood.



\section{Methods}
\label{sec:methods}

\subsection{Methods selection}
\label{ssec:methods:selection}


Several sub-topics, including anomaly detection (AD), novelty detection, uncertainty estimation (UE), and outlier detection, share motivation and methodology with OOD detection. Despite subtle differences between these topics, the approaches are similar, and most of them can be applied to OOD detection with minimal changes, as shown in \cite{yang2022openood}. So we follow the structure of \cite{yang2022openood} and select the core methods from OOD detection, UE, and AD. In our selection, we prioritize the methods already implemented for medical imaging, e.g., in \cite{karimi2022improving}, \cite{jungo2019assessing}, and \cite{zimmerer2022mood}.

As a universal baseline, the maximum probability of softmax outputs can be used to detect OOD samples without any model modifications \cite{hendrycks2016baseline}. In practice, however, the entropy of the same softmax outputs (\textbf{Entropy}) is used instead \cite{jungo2019assessing,karimi2022improving,Mehrtash_2020}. We consider Entropy a starting point for all other approaches and show its performance in our task.

The softmax entropy captures the total uncertainty, while the OOD measure corresponds only to the epistemic uncertainty, as explained in \cite{smith2018understanding}. Thereby, one can use epistemic uncertainty estimation to improve over Entropy. Among the others, Deep \textbf{Ensemble} \cite{lakshminarayanan2017simple} is considered the state-of-the-art approach to UE. To use Ensemble, one computes mutual information or variance over several predictions for a single image to obtain epistemic uncertainty map. An alternative way to obtain multiple predictions is Monte-Carlo dropout (\textbf{MCD}) \cite{gal2016dropout}, which we also include in our comparison.  

Further, we include the approach of \cite{karimi2022improving}, which directly addresses OOD detection on 3D medical images. The authors apply singular value decomposition (\textbf{SVD}) to the network features and use singular values as an image embedding. OOD score is calculated as the distance from a sample's embedding to its nearest neighbor from a training set. 


A better uncertainty estimation can be obtained by modifying the downstream model, although such modifications can harm the model's performance. We include one such popular modification, generalized ODIN (\textbf{G-ODIN}) \cite{hsu2020generalized}, in our study. Finally, OOD scores can be obtained with an auxiliary model, solely dedicated to detecting anomalies in data. Such AD methods were extensively compared in the Medical Out-of-Distribution (MOOD) challenge \cite{david_zimmerer_2022_6362313}. We choose the best solution from MOOD 2022, implement, and include it in our experiments under the name \textbf{MOOD-1}.

Discussing the auxiliary AD models, we intentionally exclude the reconstruction-based methods (e.g., auto-encoders, generative-adversarial nets) from our consideration. Firstly, these methods performed substantially worse in MOOD 2022 than self-supervised learning-based ones (e.g., MOOD-1) \cite{zimmerer2022mood}. \cite{liang2022omni} also showed that this type of methods scores far behind self-supervised learning. And \cite{meissen2022pitfalls} highlighted the severe limitations of auto-encoders applied to OOD detection in a similar setup. Given this critique, we do not include the reconstruction-based approaches in our experiments.

So, we consider the following methods: Entropy, Ensemble, MCD, SVD, G-ODIN, and MOOD-1. Since some of them are designed for the downstream classification task, we detail their adaptation to segmentation below. 

\subsection{Methods implementation}
\label{ssec:methods:implementation}

To preserve a fair comparison, we add only trivial and unavoidable modifications. We also test (in preliminary experiments) any additional component or a critical hyperparameter of every method and select the best performing setting.


\subsubsection{Entropy} Our downstream task is binary segmentation, where the sigmoid function is applied to the network's outputs. We note that two-classes softmax can be derived from the sigmoid. Then, Entropy follows the implementation from \cite{Mehrtash_2020} and \cite{karimi2022improving}, computing the average entropy value over the predicted area (i.e., positive class). We set OOD score to 0 in the case of empty predicted mask.

\subsubsection{Ensemble} We train 5 U-Net models with different initializations and calculate the uncertainty map as the voxel-wise standard deviation of the five corresponding predictions. OOD score is the average of this uncertainty map.

\subsubsection{MCD} We implement MCD by introducing a dropout layer before every down- and up-sampling layer in U-Net. To obtain an uncertainty map, we calculate voxel-wise standard deviations of 5 inference steps with a dropout rate of 0.1. OOD score is the average of this uncertainty map.

\subsubsection{SVD} We follow \cite{karimi2022improving} without any changes.

\subsubsection{G-ODIN} We preserve the original structure of the G-ODIN output layer \cite{hsu2020generalized}; the only difference is that we substitute the linear layers with the convolution ones. These convolution layers have kernels of size $1^3$, so the procedure remains equal to classification of every voxel. To obtain the uncertainty map, we use the best reported \texttt{G-ODIN DeConf-C$^*$} variant. 

In Ensemble, MCD, and G-ODIN, using mutual information or entropy and averaging the uncertainty map over only the predicted area, as in Entropy, harms the performance. Thus, we use simple averaging of uncertainty. 

\subsubsection{MOOD-1} The top-performing MOOD solutions generate synthetic anomalies and train a network to segment them \cite{zimmerer2022mood}. So our MOOD-1 implementation is based on this cut-paste-segment approach, which won MOOD 2021 \cite{cho2021self}. We then supplement it with technical improvements from the 2022's best solution, such as one-cycle learning and ensembling over 5 models. The subject-level OOD score is calculated as the mean of the top-100 anomaly probabilities.

\subsubsection{Volume predictor} To demonstrate that some semantic differences might be trivial from the model's perspective but not captured by other methods, we use the total volume of prediction (positive class) as an OOD score. Since a predicted volume can vary in any direction, we consider sample an outlier if the volume is below $\frac{q}{2}$-th or above $100 - \frac{q}{2}$-th percentile of the ID, thus retaining $100 - q$ TPR. 

\subsection{Intensity Histogram Features}
\label{ssec:methods:ihf}

\begin{figure*}[h]
    \begin{center}
        \includegraphics[width=\textwidth]{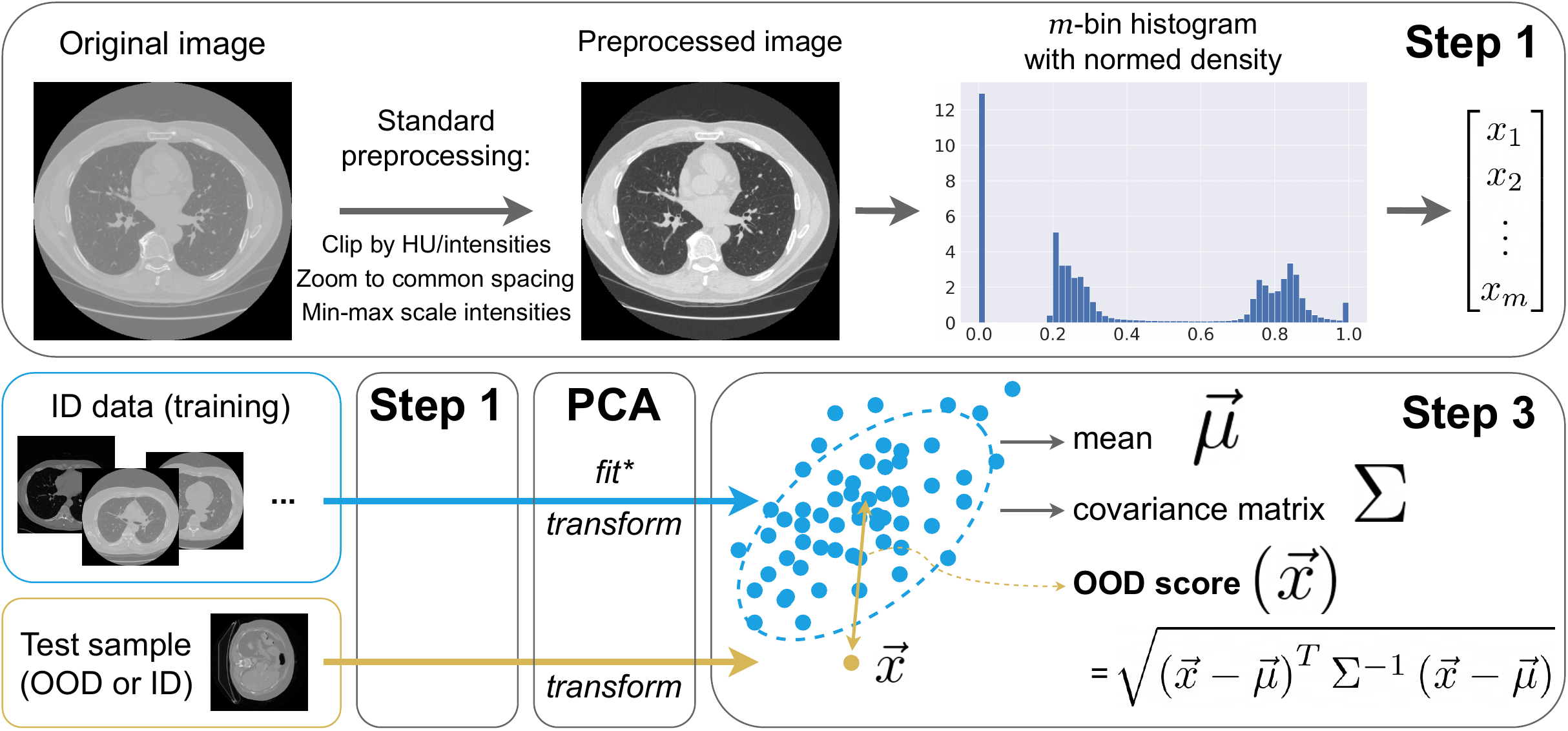}
        \caption{The proposed OOD detection method, called \textit{Intensity Histogram Features (IHF)}. It consists of three steps: calculating a $m$-dimensional vector as a histogram bin values from the preprocessed image (\textit{Step 1}), fitting and applying \textit{PCA} to the occuring data, and calculating Mahalanobis distance between a test vector and ID samples distribution (\textit{Step 3}). We apply IHF to the 3D images and illustrate the process using 2D axial slices for simplicity. (*PCA is fitted once on all training data.)}
        \label{fig:method}
    \end{center}
\end{figure*}

To contest the DL algorithms, we propose an unsupervised method based on image intensity histograms as embeddings. Our design is motivated by two other works. Firstly, \cite{karimi2022improving} showed that SVD can efficiently reduce full-image-sized network features. We note a space for improvement in their method -- one can optimize the choice of the network's layer to apply SVD. Here, \cite{zakazov2021anatomy} suggested that the earlier network's layers contain the most domain-specific information. Following the latter suggestion, we hypothesize that we can extract enough domain-specific information directly from the image (i.e., the zeroth network's layer). A histogram is a convenient way to do so.

We present our method, called Intensity Histogram Features (IHF), schematically in Fig.~\ref{fig:method}. It consists of three steps: (1) calculating intensity histograms of images and using them as vectors, (2) reducing their dimensionality with PCA, and (3) running the outlier detection algorithms on these vectors.

\paragraph*{Step 1: preprocessing and histograms} All images undergo the same preprocessing pipeline to standardize the intensity distribution:

\begin{enumerate}
    
    \item We interpolate images to the same spacing. So in all CT and MRI experiments, we use $1 \times 1 \times 1.5$ mm.

    \item We clip image intensities to $[-1350; 300]$ Hounsfield units for CT (a standard lung window) and $[1^{th}$percentile; $99^{th}$percentile$]$ for MRI.

    \item We min-max-scale image intensities to the $[0, 1]$ range.
    
\end{enumerate}

Given a preprocessed image $x$, we then compute a probability density function of its intensities in $m$ bins, a histogram $e(x) \in \mathbb{R}^m$, and further use these vectors $e(x)$.

\paragraph*{Step 2: Principal Component Analysis (PCA)} As an optional step, we use PCA to reduce the dimensions $m$. The main reason to use it is that some outlier detection algorithms at \textit{Step 3} behave unstable in high dimensional spaces. For instance, calculating Mahalanobis distance require reversing the empirical sample covariance matrix, and this matrix is likely to become ill-conditioned or singular with larger $m$.

Therefore, we fit PCA$_v$ once on the training data $E_{tr}$ to preserve $v = 99.99\%$ of the explained variance. This mostly eliminates the potential instability while preserving the distribution properties. $E_{tr}$ consists of row-vectors $e(x_{tr})$ for all training images $x_{tr} \in X_{tr}$. Further, we use transformed vectors $\tilde{e}(x) = \text{PCA}_v (e(x))$. 

\paragraph*{Step 3: OOD detection algorithm}

To calculate the OOD score for $x$, we can apply any distance- or density-based outlier detection method. As in \cite{lee2018simple}, we can calculate Mahalanobis distance $S_{Mah}(x)$: 

\begin{equation}
    \label{eq:mah}
    S_{Mah}(x) = \sqrt{ \left( \tilde{e}(x) - \hat{\mu} \right)^T \hat{\Sigma}^{-1} \left( \tilde{e}(x) - \hat{\mu} \right) },
\end{equation}

\noindent
where $\hat{\mu}$ and $\hat{\Sigma}$ are the estimated mean and covariance matrix on the training set, $\hat{\mu} = \frac{1}{|X_{tr}|} \sum_{x_{tr} \in X_{tr}} \tilde{e} \left(x_{tr}\right)$ and $\hat{\Sigma} = \frac{1}{|X_{tr}|} \sum_{x_{tr} \in X_{tr}} \left( \tilde{e} (x_{tr}) - \hat{\mu} \right) \left( \tilde{e} (x_{tr}) - \hat{\mu} \right)^T$. 
Alternatively, one can calculate the distance to the nearest neighbor (min-distance) $S_{NN}(x)$, as in \cite{karimi2022improving}:

\begin{equation}
    \label{eq:nn}
    S_{NN}(x) = \min_{x_{tr} \in X_{tr}} || \tilde{e} (x) - \tilde{e} (x_{tr}) ||_2.
\end{equation}

Using $S_{Mah}$ (Eq.~\ref{eq:mah}) and $S_{NN}$ (Eq.~\ref{eq:nn}) corresponds to the methods called IHF-Mah and IHF-NN, respectively. We include them in comparison and ablation study independently.

\section{Experimental setup}
\label{sec:exp}

\subsection{Downstream task}
\label{sssec:data:exp:segm}

We have 3D CT and MRI datasets with a binary segmentation task. So we adhere the standard approaches to train a segmentation model in all methods that require the latter.

\subsubsection*{Data preprocessing} We describe preprocessing in IHF, Step 1 (Sec.~\ref{ssec:methods:ihf}); it is the same in all experiments and it is the minimum allowing the correct DL model training.

\subsubsection*{Architecture and training} In all experiments, we use 3D U-Net \cite{isensee2018no}, a standard architecture for segmentation. We train it on patches of $64$ axial slices, with a batch size of $3$, Adam optimizer, and learning rate of $10^{-4}$ for $30$ epochs, $1000$ iterations each. In a batch, patches from different images are padded if necessary. We minimize the sum of Binary Cross-Entropy and Focal Tversky losses \cite{abraham2019novel} to achieve high segmentation sensitivity.

\paragraph*{Segmentation evaluation} We train all models on the training part of ID datasets. Then, we can evaluate their segmentation quality on the corresponding testing part of the OOD datasets, showing its possible decline. These segmentation results are given in Tab.~\ref{tab:segm_ct} for the CT and in Tab.~\ref{tab:segm_mri} for the MRI datasets.

\subsection{OOD detection evaluation}
\label{sssec:data:exp:ood}

Given the testing part of the ID dataset, we measure the OOD detection quality against it for all the suggested OOD setups, similarly to the classification task. Outliers occur rarely in practice, so we aim to measure detection quality when most of the ID samples are being preserved w.r.t. relatively rare OOD events. In this case, one of the most convenient classification metrics to use is false positive rate at 95\% true positive rate (FPR), so we consider FPR our primary metric Nonetheless, for the consistency with other studies, we report AUROC in the supplementary materials.

\section{Results}
\label{sec:results}

In this section, we report on our experiments and results. We start by benchmarking all considered methods, then present the analysis of the benchmark design and finish with the ablation study of the methods on synthetic data.

\begin{table*}[h]
\centering
\caption{Comparison of the considered OOD detection methods in terms of FPR@TPR95\% scores (lower is better). We highlight the best scores in every row in \textbf{bold} and ranked the methods by their average performance. The first and second sections correspond to CT and MRI setups, respectively.}
\resizebox{\textwidth}{!}{%
\begin{tabular}{llllllllll}
\toprule
OOD setup &            IHF-NN &             SVD &             IHF-Mah &      MOOD-1 &           G-ODIN &             Volume &   MCD & Ensemble &   Entropy \\
\midrule
Location (Head)           &  \textbf{.00}  &  \textbf{.00}  &  \textbf{.00}  &            .12 &            .55 &            .53 &  .36 &     .51 &  .56 \\
Location (Liver)          &            .51 &  \textbf{.13}  &            .64 &            .56 &            .56 &            .84 &  .89 &     .93 &  .78 \\
Population (COVID-19)     &            .54 &            .75 &            .72 &  \textbf{.51}  &            .54 &            .82 &  .58 &     .58 &  .87 \\
Scanner                 &            .88 &            .89 &            .85 &  \textbf{.73}  &            .92 &            .86 &  .89 &     .90 &  .83 \\
Synthetic (Elastic)       &  \textbf{.15}  &            .37 &            .67 &            .16 &            .59 &            .81 &  .42 &     .37 &  .84 \\
Synthetic (Image noise)  &            .49 &            .37 &            .62 &  \textbf{.11}  &            .89 &            .85 &  .87 &     .82 &  .81 \\
\midrule
Population (Glioblastoma) &  \textbf{.00}  &  \textbf{.00}  &  \textbf{.00}  &            .10 &            .21 &            .01 &  .85 &     .81 &  .86 \\
Population (Healthy)      &  \textbf{.00}  &  \textbf{.00}  &  \textbf{.00}  &            .11 &  \textbf{.00}  &  \textbf{.00}  &  .88 &     1.0 &  .85 \\
Scanner                   &  \textbf{.00}  &  \textbf{.00}  &  \textbf{.00}  &            .15 &  \textbf{.00}  &            .74 &  .63 &     .66 &  .89 \\
Synthetic (K-space noise) &  \textbf{.00}  &            .36 &  \textbf{.00}  &            .88 &            .88 &            .90 &  .82 &     .77 &  .73 \\
Synthetic (Anisotropy)    &            .09 &            .20 &  \textbf{.05}  &            .57 &            .88 &            .93 &  .77 &     .77 &  .81 \\
Synthetic (Motion)        &  \textbf{.00}  &            .58 &  \textbf{.00}  &            .73 &            .93 &            .94 &  .85 &     .88 &  .91 \\
Synthetic (Image noise)   &            .47 &            .33 &            .47 &  \textbf{.30}  &            .56 &            .71 &  .78 &     .75 &  .75 \\
\midrule
CT average                &            .43 &            .42 &            .58 &  \textbf{.36}  &            .67 &            .79 &  .67 &     .68 &  .78 \\
MRI average               &            .08 &            .21 &  \textbf{.07}  &            .41 &            .50 &            .60 &  .80 &     .81 &  .83 \\
\bottomrule
\end{tabular}}
\label{tab:res_fpr}
\end{table*}

\subsection{Benchmarking}
\label{ssec:results:benchmark}

Tab.~\ref{tab:res_fpr} presents the primary results of our study. Uncertainty-based methods, the ones that are not designed for segmentation, mostly fail in the suggested challenges. Entropy, Ensemble, MCD, and G-ODIN gives substantially worse (higher) FPR than the other methods, with only G-ODIN slightly surpassing a simple \textit{Volume} predictor. Methods dedicated to segmentation perform better on average. For instance, our implementation of the best MOOD 2022 solution, MOOD-1, achieves $0.36$ and $0.41$ average FPR on CT and MRI data, respectively. SVD improves further; it appears the only reliable studied method, providing $0.42$ and $0.21$ mean FPR.

Then, SVD performance is contested by the proposed IHF. In a combination with min-distance, IHF-NN provides the best average score across studied challenges: 0.43 and 0.08 FPR, respectively. In a combination with Mahalanobis distance, IHF-Mah provides practically worse results in the CT setups. Although IHF-Mah performs weaker, it was historically the first and we submitted it to MOOD 2022 ($m=150$, no PCA). We placed second (team AIRI\footnote{\url{http://medicalood.dkfz.de/web/}}) with the earliest IHF version, supporting its robustness by the independent evaluation.

\begin{figure*}[h]
    \begin{center}
        \includegraphics[width=\textwidth]{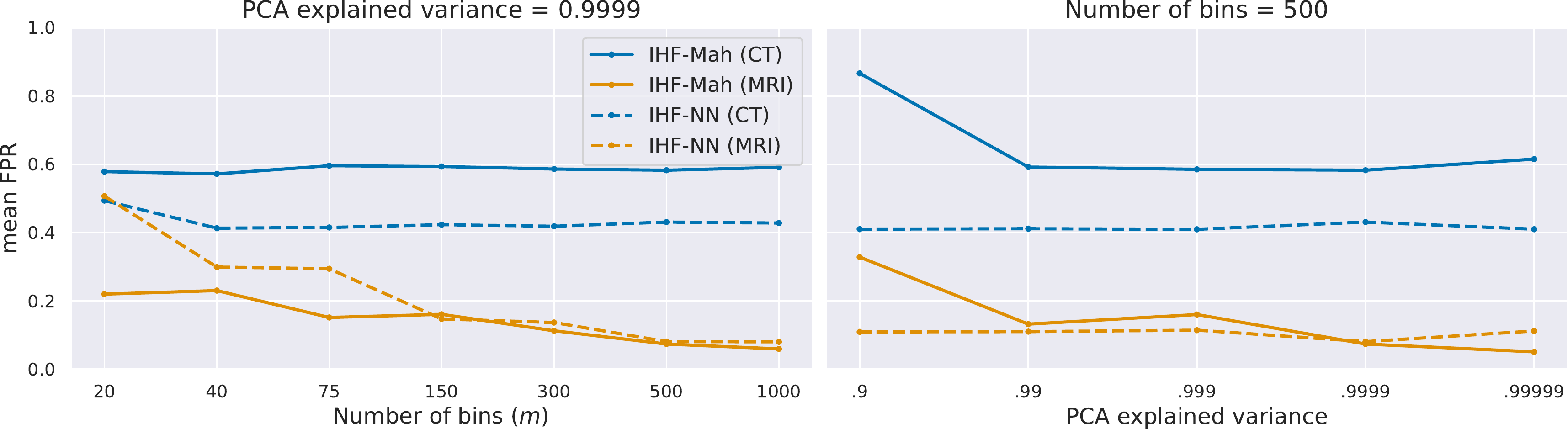}
        \caption{Dependence of IHF on its two hyperparameters: the number of histogram bins ($m$) and explained variance in PCA ($v$). We give the results for both IHF variants and CT and MRI setups.}
        \label{fig:ihf_hyp}
    \end{center}
\end{figure*}

We also conduct an ablation study to verify IHF robustness. As shown in Fig.~\ref{fig:ihf_hyp}, we test IHF performance varying its only two parameters, the number of bins ($m$) and explained variance ratio ($v$). Our findings indicate a consistent behavior and numerical stability regardless of the parameter choice, with a slight trend of improved quality at a larger $m$.

Both IHF variants perform comparable or better on average than SVD and, consequently, the other studied methods. Therefore, we conclude that the histograms of image intensities are descriptive enough to detect most of the suggested OOD cases, while neural networks might omit important domain features that could be used in this problem. We thus hypothesize that neural networks-based OOD detection can be further improved and leave this promising direction for future research.

We present the same comparison in terms of AUROC in Tab. \ref{tab:res_auroc}. Although AUROC is not our primary metric, it roughly preserves the same relative ranking of the studied methods, not contradicting to our main message.

\subsection{In-depth benchmark analysis}
\label{ssec:results:philosophy}


Further, we emphasize the significance of constructing a \textit{correct} benchmark to study the methods. Analysis of our experimental results suggests the following:

\begin{itemize}
    
    \item The benchmark should contain diverse OOD challenges.

    \item Challenges should represent clinically occurring cases.

    \item Potential biases in the benchmark should be explored using simple methods, such as IHF or Volume predictor.

\end{itemize}

It is often possible to develop a method tailored to specific OOD sources where it thrives but fails in the other setups. For example, G-ODIN demonstrates near-perfect results in the Population and Scanner setups on MRI data but yields the worst scores in the others. In practice however, the precise anomaly source is always unknown, and a general OOD detector, with an acceptable average performance, is needed. The true method effectiveness can be estimated only in the context of diverse challenges.

Secondly, OOD sources should accurately represent or simulate the clinically occurring cases. For instance, the Synthetic (Noise) setup, as introduced in \cite{zimmerer2022mood} and reproduced in our study, is not supported by any medical imaging process. MOOD-1 achieves the highest performance in this setup because its training objective closely aligns with the anomaly synthesis process. However, performing well in this and similar cases is of no clinical value and, consequently, biases the methods' evaluation towards explicitly unrealistic scenarios.

Finally, our analysis reveals that OOD challenges might contain implicit but trivial features. If a benchmark focuses solely on any such feature, we can design a method that exploits this feature, leading to deceptive conclusions about the generalized performance. Instead, we suggest using such methods to reveal biased features beforehand. For example, near-perfect IHF results in several setups demonstrate that certain anomalies are actually trivial intensity changes, reinforcing the need of designing diverse benchmarks.

\begin{table}[h]
\centering
\caption{Fechner correlation coefficients between IHF-NN, Volume and the other studied methods' performance.} 
\resizebox{0.5\textwidth}{!}{%
\begin{tabular}{llllllllll}
\toprule
 &  IHF-NN & SVD & IHF-M & MOOD & GODIN & Vol & MCD & Ens & Ent \\
\midrule
IHF-NN           &  1.00 & 0.38 & \textbf{0.85} & 0.23 & -0.08 & 0.23 & 0.08 & -0.08 & -0.23 \\
Volume          &   0.23 & \textbf{0.54} & 0.38 & 0.38 & 0.38 & 1.00 & 0.23 & 0.08 & 0.23  \\

\end{tabular}}
\label{tab:res_corr}
\end{table}

To ensure methods' generalization, we calculate the Fechner correlation between their results and results of the \textit{simple methods}. We show that, apart from SVD, the other methods do not exhibit a strong correlation with the Volume or IHF scores (Tab.~\ref{tab:res_corr}). So the examined methods mostly do not rely on trivial features, such as image intensity distribution. However, SVD shows a correlation of 0.54 with the Volume scores, suggesting its hindered generalization on new sources of data with small difference in the predicted area volume. 

\begin{figure*}[h]
    \begin{center}
        \includegraphics[width=\textwidth]{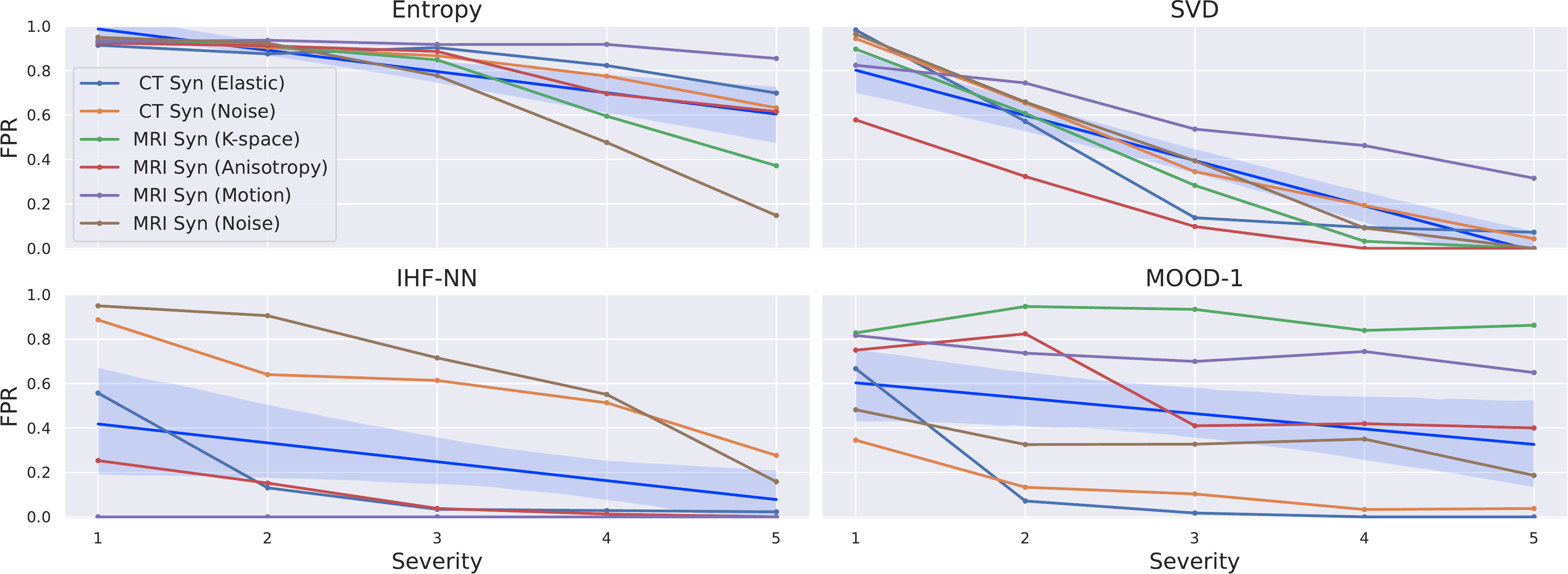}
        \caption{FPR under synthetically distorted data for every distortion severity level. Blue line indicates method's average trend across presented challenges with 95\% confidence interval. The other UE methods (MCD, Ensemble, and G-ODIN) are excluded since their average trend is similar to Entropy.}
        \label{fig:fpr_sev}
    \end{center}
\end{figure*}

\subsection{Ablation study on synthetic data}
\label{ssec:results:synthetic}

In Fig.~\ref{fig:fpr_sev}, we show the OOD detection results on synthetic data for different levels of distortion. A larger severity level indicates an image corruption with the larger magnitude, where the magnitudes are chosen perceptually from 1 (barely noticeable distortion) to 5 (heavily distorted image). The general trend is that more distorted images are easier to detect. Here, SVD exhibits the steepest average slope and behaves almost linearly with the increasing severity level, suggesting that we have considered challenging but solvable tasks.

Different methods exhibit different sensitivity to the level of distortion required for detection. Entropy and the other UE methods begin to operate effectively only at level 3, while IHF can detect anomalies at the minimal level. So we conclude that the methods should be studied across a wide range of the anomaly severity levels. Additionally, we show that $\text{MOOD-1}$ depends more on the OOD source than severity level: it completely fails in Motion and K-space setups, while almost perfectly detects Noise and Elastic deformations independently of the severity level. Moreover, MOOD-1 and IHF behave inversely to each other in Noise and Elastic setups. Such diverse behavior suggests the need to study methods also across a wide range of anomaly types.


\section{Discussion}
\label{sec:discussion}

Besides benchmarking the OOD detection methods, our study also suggests practical ideas on building a correct benchmark. We mainly highlight the data diversity in multiple dimensions and clinical relevance of the setups. However, we leave several critical questions undeveloped, thus opening the corresponding research directions, which we discuss below.




%

\subsubsection{Benchmark design}

We select the datasets that represent the clinically occurring sources of OOD data. And we confirm the importance of the constructed challenges by the degraded downstream performance (Tab.~\ref{tab:segm_ct}, Tab.~\ref{tab:segm_mri}). However, we cannot state with certainty that the highlighted sources are the only differences between the distributions. While we name the primary difference in each case (e.g., acquisition parameters, patient populations), the distributions might differ due to several causes and the other causes exist outside of our consideration. So the development of a more refined benchmark with the controlled distinction between the OOD sources would greatly facilitate further research in this area.

Furthermore, the question of whether any semantic difference should be considered abnormal requires additional investigation. For instance, Population (Healthy) setup is considered OOD due to this apparent semantic difference -- healthy cases instead of pathology ones. Segmentation models often yield correct empty predictions for such images regardless any OOD detection decision. Rejecting a correct prediction in this case should be considered a false negative instead of true positive outcome, lowering the false positive rate.


\subsubsection{Uncertainty estimation} Our study verifies that the epistemic uncertainty is better suited for OOD detection than the total uncertainty, since MCD and Ensemble achieve better results than Entropy. Nevertheless, a question of how to aggregate the uncertainty map into a single score remains open. On the one hand, aggregating uncertainty over the predicted volume offers certain advantages compared to the whole volume, especially when dealing with 3D images, where the area of interest may occupy only a small portion of the entire image. While this aggregation shows improved results for the Entropy method, it cannot rank samples with an empty predicted mask and does not trigger on anomalies outside of the predicted area. Contrary, aggregating uncertainty as a simple average over the whole image provides better results for MCD and Ensemble. Developing a reasonable UE method for 3D images, thereby, is a possible direction for future research.

\subsubsection{Other IHF applications}

Additionally, we explored alternative applications of the proposed IHF method. We noted its strong performance in such medical imaging tasks as contrast detection and domain classification. In this paper, however, we do not delve into the potential IHF applications, as they lie beyond the scope of the OOD detection problem.

\section{Conclusion}
\label{sec:conclusion}

In this paper, we have conducted an extensive investigation of OOD detection on 3D medical images. Our results revealed that the established approaches, including uncertainty estimation and anomaly detection ones, are unable to provide reliable performance. They provide unacceptably high number of false positives (0.31 mean FPR at best) and fail to generalize. They can also be improved. To demonstrate this, we developed a histogram-based method, IHF, that achieves comparable and often superior results to its competitors. Thereby, we indicated that the distribution shifts in 3D medical imaging can often be detected using intensity histograms, while the DL algorithms neglect this domain feature. Although IHF achieves better average results, its performance is surpassed in multiple challenges, emphasizing the need and possibility for developing a robust and general OOD detection method.

We constructed and released the corresponding challenges as a benchmark for OOD detection on 3D medical images, proposing IHF as a solid baseline to contest new methods. 


\paragraph*{Acknowledgments} We acknowledge the National Cancer Institute and the Foundation for the National Institutes of Health, and their critical role in the creation of the free publicly available LIDC/IDRI Database used in this study.

\appendices

\bibliographystyle{IEEEtran}
\bibliography{bibliography/1_intro,bibliography/2_related,bibliography/3_methods,bibliography/4_experiments}

\newpage
\newpage
\onecolumn

\section{Segmentation quality on ID and OOD datasets}
\label{appendix:segm}


\begin{table*}[h]
    \centering
    \caption{Segmentation quality for all considered CT setups. We use Dice score (DSC, higher is better) and average number of False Positives per image (Avg. FP, lower is better) as our metrics.}
    \resizebox{\textwidth}{!}{%
    \begin{tabular}{l c c c c c c c}
        \toprule
        \multirow{2}{*}[-0.2em]{Metric} & ID dataset & \multicolumn{6}{c}{OOD challenges} \\
        \cmidrule{2-8}
        
        & LIDC & Loc(Head) & Loc(Liver) & Scanner & Pop(COVID) & Syn(Noise) & Syn(Elastic) \\

        \midrule
        DSC $(\uparrow)$ & $.50 \pm .27$ & $.21 \pm .41$ & $.03 \pm .16$ & $.18 \pm .24$ & $.00 \pm .00$ & $.32 \pm .30$ & $.13 \pm .23$ \\

        Avg. FP $(\downarrow)$ & $5.2 \pm 6.9$ & $2.6 \pm 4.6$ &  $10 \pm 16$ & $ 6 \pm 15$ & $13 \pm 10 $ & $12\pm 25$ & $5.4 \pm 4.5$ \\

    \bottomrule
    \end{tabular}}
    \label{tab:segm_ct}
\end{table*}

\begin{table*}[h]
    \centering
    \caption{Segmentation quality for all considered MRI setups in terms of Dice score (DSC).}
    \resizebox{\textwidth}{!}{%
    \begin{tabular}{l c c c c c c c c}
        \toprule
        \multirow{2}{*}[-0.2em]{Metric} & ID dataset & \multicolumn{6}{c}{OOD challenges} \\
        \cmidrule{2-9}
        
        & VS-Seg & Pop(Healthy) & Pop(Glioma) & Scanner & Syn(K-space)  & Syn(Anisotropy)  & Syn(Motion) & Syn(Noise) \\
        
        \midrule
        DSC $(\uparrow)$ & $.92 \pm .05$ & $.85 \pm .36$ & $.86 \pm .34$ & $.89 \pm .12$ & $.83 \pm .023$ & $.87 \pm .01$ & $.91 \pm .06$ & $.57 \pm .41$ \\

    \bottomrule
    \end{tabular}}
    \label{tab:segm_mri}
\end{table*}

\section{OOD detection results in terms of AUROC}
\label{appendix:auroc}

\begin{table*}[h]
    \centering
    \caption{Comparison of the considered OOD detection methods in terms of AUROC scores (higher is better). We highlight the best scores in every row in \textbf{bold} and ranked the methods by their average performance. The first and second sections correspond to CT and MRI setups, respectively.}
    \resizebox{\textwidth}{!}{%
    \begin{tabular}{llllllllll}
\toprule
OOD setup &            IHF-NN &             IHF-Mah &             SVD &           G-ODIN &   Volume & MOOD-1 &   MCD & Ensemble &   Entropy \\
\midrule
Location (Head)           &  \textbf{1.0}  &  \textbf{1.0}  &  \textbf{1.0}  &            .83 &  .73 &       .83 &  .85 &     .79 &  .62 \\
Location (Liver)          &            .89 &            .85 &  \textbf{.97}  &            .88 &  .65 &       .61 &  .42 &     .45 &  .67 \\
Population (COVID-19)     &  \textbf{.88}  &            .83 &            .74 &            .86 &  .76 &       .66 &  .79 &     .80 &  .72 \\
Scanner                 &  \textbf{.73}  &  \textbf{.73}  &            .58 &            .72 &  .68 &       .51 &  .58 &     .55 &  .65 \\
Synthetic (Elastic)       &  \textbf{.97}  &            .83 &            .86 &            .85 &  .77 &       .78 &  .84 &     .85 &  .65 \\
Synthetic (Image noise)  &            .81 &            .77 &  \textbf{.84}  &            .75 &  .67 &       .80 &  .56 &     .61 &  .59 \\
\midrule
Population (Glioblastoma) &  \textbf{1.0}  &  \textbf{1.0}  &  \textbf{1.0}  &            .96 &  .68 &       .87 &  .44 &     .41 &  .14 \\
Population (Healthy)      &  \textbf{1.0}  &  \textbf{1.0}  &  \textbf{1.0}  &  \textbf{1.0}  &  .68 &       .86 &  .44 &     .16 &  .15 \\
Scanner                   &  \textbf{1.0}  &  \textbf{1.0}  &  \textbf{1.0}  &  \textbf{1.0}  &  .77 &       .83 &  .70 &     .74 &  .59 \\
Synthetic (K-space noise) &  \textbf{1.0}  &  \textbf{1.0}  &            .86 &            .81 &  .66 &       .24 &  .56 &     .63 &  .66 \\
Synthetic (Anisotropy)    &  \textbf{.98}  &  \textbf{.98}  &            .94 &            .81 &  .68 &       .57 &  .63 &     .63 &  .71 \\
Synthetic (Motion)        &            .99 &  \textbf{1.0}  &            .75 &            .78 &  .68 &       .48 &  .57 &     .54 &  .57 \\
Synthetic (Image noise)   &            .81 &            .83 &            .85 &  \textbf{.88}  &  .66 &       .77 &  .58 &     .58 &  .56 \\
\midrule
CT average                &  \textbf{.88}  &            .84 &            .83 &            .82 &  .71 &       .70 &  .67 &     .68 &  .65 \\
MRI average               &  \textbf{.97}  &  \textbf{.97}  &            .92 &            .89 &  .69 &       .66 &  .56 &     .53 &  .48 \\
\bottomrule
\end{tabular}  

    }
    \label{tab:res_auroc}
\end{table*}

\end{document}